# Accelerating the Fusion Workforce

*A Report from the National Science Foundation Funded Clean Energy Technology Conference*

C. Paz-Soldan, E. Belonohy, T. Carter, L.E. Coté, E. Kostadinova, C. Lowe, S. L. Sharma, S. de Clark, J. Deshpande, K. Kelly, V. Kruse, B. Makani, D. Schaffner, and K. Thome

## Executive Summary

The fusion energy research and development landscape has seen significant advances in recent years, with important scientific and technological breakthroughs and a rapid rise of investment in the private sector. The workforce needs of the nascent fusion industry are growing at a rate that academic workforce development programs are not currently able to match.

This report presents the findings of the National Science Foundation (NSF) funded **Workforce Accelerator for Fusion Energy Development** conference held in Hampton, Virginia in May 2024. The event brought together experts from industry, academia, and the national laboratories to acknowledge the challenges facing the U.S. fusion workforce and develop recommendations to address them. A broad range of relevant topics were addressed in 26 distinct white papers[1] (which will be included as supplemental data to this report). This summary highlights these findings, which are presented in detail in the main report body text.

**Technical Areas for Fusion and Broader STEM Education**: Plentiful research and development opportunities exist from which to grow a talented and skilled fusion workforce. Several areas are broadly applicable and offer pathways to STEM education serving multiple fields, **enabling workforce mobility into and out of the fusion ecosystem**. We recommend *emphasizing unique and broadly applicable fusion enabling technologies in fusion workforce development strategies*. This report highlights which technical areas for investment are broadly applicable to fusion and which areas are broadly applicable to multiple high-technology fields.

**Educating, Training, and Developing the Fusion Workforce:** Fusion and adjacent programs at universities, together with broad skill development at community colleges and trade schools, are an essential element of our efforts to accelerate the fusion workforce. Academic programs concentrate talent, connect to their local communities, and foster the interdisciplinary work needed to technically advance fusion. Indeed, today's fusion industry in large part spun-out from the academic sector. Despite this inherent strength and advantages for workforce development, we find academic institutions are presently under-utilized and significant scope exists to amplify the efforts of this sector. Specifically, we propose targeted support to *develop a comprehensive K-12 public engagement and outreach strategy,* as well as *establish subject matter expert mobility schemes*, and finally *expand continuing education, fellowship, and internship opportunities*.

---

[1] White papers generated for the NSF-CET meeting will be marked "WP" when referenced in the text (e.g., Keane et al., 2024 WP).

***The fusion curriculum*** was surveyed as part of this effort. There exists a burgeoning interest among students at all levels in fusion-related work. However, a significant gap in skills persists, necessitating targeted training programs. To bridge this skills gap, we propose governmental agencies ***develop and implement a national strategy for fusion curriculum*** as well as ***launch programs to catalyze faculty positions in fusion engineering and technology***.

***Academic sector participants*** also must target their educational offerings to best leverage the expanded interest in fusion. Academic institutions should: *Improve accessibility to education by developing shorter certificate programs and publicly available courses* as well as *support the production of easily accessible, engaging, interactive and inclusive educational materials*. Broad participation would also be improved by *encouraging open-source ecosystem development for fusion hardware and software*.

**Domestic and International Partnerships:** Fusion energy development benefits from an existing ecosystem of performers. Establishing synergistic partnerships between these entities will be essential to accelerate the development of the needed workforce. We propose that both the community and agencies ***develop partnerships with existing public laboratories, private sector and other relevant tech sectors*** as well as ***establish collaborative projects between US institutions and international partners on education and workforce development***.

**Workforce Scaling and Challenges:** This report provides projections showing that the private sector fusion workforce is expected to grow substantially. We find that academia is struggling to keep up with this growth. As a starting point, what this workforce will or should look like lacks certainty. We make recommendations to evaluate the distribution, needs, and challenges of the current and future US fusion workforce, keeping in mind the level of accessibility and diversity in the workforce. We recommend a ***formal study evaluating the distribution, needs and challenges of the current and future US fusion workforce*** as well as support for efforts to ***document the level and trajectory of accessibility and diversity in the fusion workforce***.

**Community Benefits and Inclusion:** We conclude our review with the most important recommendations to foster broad community benefits and inclusion as fusion energy is developed and commercialized. A new clean energy technology offers the opportunity to confront and avoid issues that plagued past deployments and position community benefits at the center of our development. The deployment of a new energy source also offers the opportunity to foster a diverse and talented workforce. The fusion community as a whole must work tirelessly to *strengthen the fusion community through collaboration, equity, and inclusion* as well as *support efforts that improve working conditions for the existing fusion workforce*. Agencies have a role to ***establish community trust and public engagement of fusion energy and its benefits using Community Benefit Planning***.

To conclude, there is a significant societal interest in the fusion field as a result of recent breakthroughs, and the base of engaged talent exists to grow the clean energy technology workforce of tomorrow. Our central recommendation is thus as follows: **Now is the time to accelerate fusion workforce development efforts through targeted agency investments.**

# Introduction

The fusion energy research and development landscape has seen significant changes in the last several years. The urgency to transition to clean energy and the advancements in fusion research have spurred a substantial increase in private investment, the establishment of new private companies, the formation of public-private partnerships, and the close-involvement of the industry in large-scale construction projects. Private investment in fusion energy has surpassed $7 billion worldwide, about 80% of which is invested into US companies, with over 45 private companies complementing the established publicly funded research centers (FIA, 2024). The industry anticipates the creation of thousands of new jobs, emphasizing the critical need for a diverse and well-trained workforce to achieve the societal goal of rapid fusion energy commercialization (The White House, 2022).

## Workforce Challenges for Fusion

The fusion energy sector requires significant workforce expansion to scale fusion as an energy source to meet the net-zero by 2050 targets. Major challenges facing the fusion energy workforce include:

1. **Workforce shortage and retention**: US National Labs, User Facilities, and academic fusion experiments are experiencing significant shortages of trained personnel at all levels, exacerbated by retention challenges. Private fusion companies are experiencing challenges in recruiting and retaining individuals with needed skills and face competition both from within and outside of the fusion sector.
2. **Knowledge retention**: Difficulties in recruiting and retaining skilled workers in the public program are compounded by knowledge gaps due to retirement of experienced staff and insufficient documentation of tacit knowledge, lessons learned and best practices from decades of hands-on experience.
3. **Education and training gaps**: There is insufficient interdisciplinary training covering various technical aspects of fusion energy at educational institutions, in vocational training programs, and workplaces as part of continuing education. Exposure to fusion in early education and hands-on learning opportunities on fusion experiments are lacking. Fusion education is not widely administered at technical and trade schools, and is available in limited capacity at undergraduate and graduate levels. Additionally, there exists a need for improved training resources available for all workforce segments for career progression.
4. **Limited support for workforce development**: The funding environment for workforce development activities is very constrained, presently supporting almost no dedicated professionals who can advance these issues or support the aforementioned training activities. Infrastructure development at academic institutions has been underfunded for decades in the fusion domain, with consequent drops in student engagement.
5. **Limited fusion-enabling technology programs**: Educational and research programs in fusion-enabling technologies are limited in terms of the different technologies supported and the number of educational institutions offering them.

6. **Public engagement and community benefit gaps:** To increase interest in joining the fusion workforce, public engagement in communities of interest need to be aware of and have a positive view of fusion energy. Furthermore, any industrial fusion activity (such as a new plant) is going to require a formal Community Benefit Plan ([Keane et al., 2024 WP](#)) as well as engaging with the communities in the proposed areas of development.
7. **Diversity, Equity, Inclusion, and Accessibility (DEIA) issues**: Challenges exist in attracting and integrating diverse team members and accommodating students from minority-serving institutions, smaller university programs, and primarily undergraduate education programs. The uneven progress in growth of women and minorities in STEM related occupations is well documented (Fasoli et al., 2024; Fry et al., 2021). This observation is valid for the fusion sector and anecdotally the participation by underrecognized is lower in the fusion sector than the national average for the STEM sector. Furthermore, at undergraduate levels and below, underrecognized students struggle to obtain skills essentially to pursue higher education, also referred to as "hidden curriculum" disadvantages.

## Fusion Workforce Accelerator Conference

As an initial step to meet these challenges, the ***[Fusion Workforce Accelerator](#)*** conference was held at Hampton, Virginia in May 2024 to outline a US workforce development strategy to meet the needs of current and future fusion development by bringing together experts from academia, industry, non-profits, national laboratories, and the federal government (i.e., NSF, DOE, congressional staff).

The conference brought stakeholders together to identify opportunities for partnership in fusion research and education with the goal of meeting the needs for a talented, diverse workforce. Well *over 100 participants* were engaged in the process, with *~70 on-site attendees*, including international participants (from the European Union and the United Kingdom). The group of people assembled were wide-ranging, from different industries and organizations, and largely unfamiliar to each-other prior to the event, indicating the event convened stakeholders who would not normally interact.

*Conference participants wrote* **26 white papers***,* spanning themes of: 1) technical research areas well suited to workforce expansion, 2) existing domestic and international workforce programs, 3) academic and curriculum needs to support the growing workforce, 4) community benefits and diversity. The conference agenda consisted of plenary white paper presentations by the authors and discussion sessions for collecting feedback and recommendations. Recommendations were then brought to the group via plenary sessions and via online forms. Informal discussions were also conducted throughout the meeting.

This report documents the findings and recommendations of this event, divided along the above topical sections. As shown in Fig 1, there is a strong connection between the development of the fusion workforce and the development of overall capability in high-tech and STEM sectors in the US. Technical areas that connect the two will be elaborated, as will the importance of knowledge transfer, curriculum, academia, community colleges, outreach, and K-12 education. By aligning and developing all these areas, the fusion workforce can grow to meet the challenge and broaden the overall technical capabilities of the US.

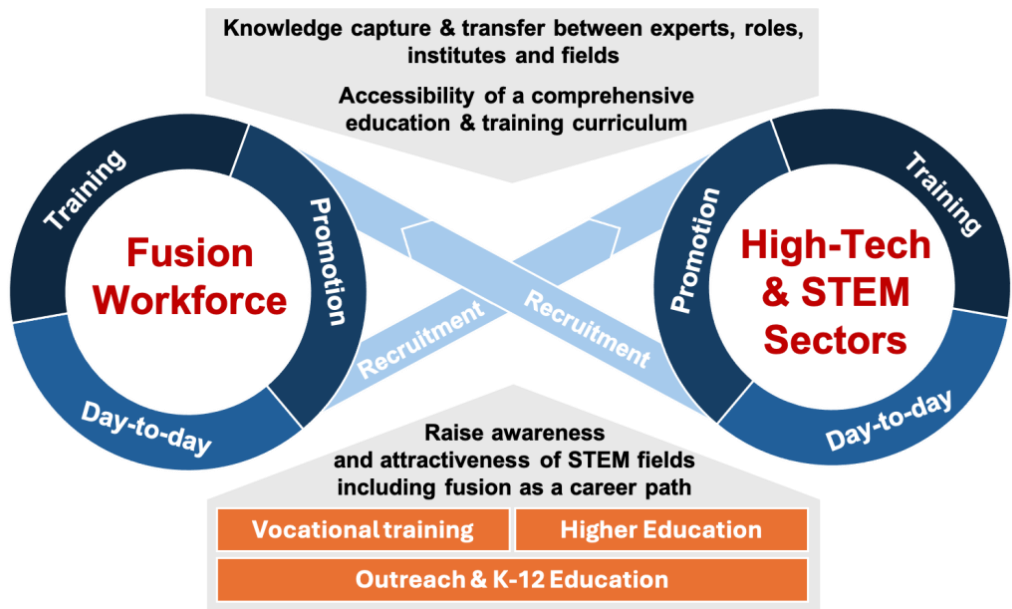

*Fig 1: Infographic detailing the interplay of the specific fusion workforce, the broader STEM workforce, and factors that improve the development and synergy of both.*

# Technical Areas for Fusion and Broader STEM Education

The fusion environment is extremely harsh in terms of temperature, particle flux, and neutron irradiation. To meet the challenge of deploying fusion energy, advancements in several technical areas are needed (Baalrud et al., 2020). Several necessary systems are presently at a low level of technical readiness, and plentiful research and development opportunities exist from which to grow a talented and skilled workforce. This section will highlight which technical areas are ripe for expanded workforce development activities, several of which are well linked to other technical areas and offer broad pathways to STEM education across multiple fields.

Historically, fusion energy education has primarily been offered through physics and nuclear engineering departments. However, Mechanical Engineering represents the top major among all hires in the private sector today, closely followed by Physics, Electrical Engineering, and Aerospace Engineering ([Deshpande & Segal, 2024 WP](#)). It is anticipated that the private sector fusion workforce will continue to draw from diverse backgrounds spanning all engineering disciplines and adjacent areas. Industry anticipates a significantly higher growth in manufacturing and operations workforce than the physics workforce. This not only signals a shift of focus from scientific exploration to commercial exploitation, but also indicates industry-driven growth in employment opportunities in fusion-enabling technologies.

**Emphasize unique and broadly applicable fusion enabling technologies in fusion workforce development strategies**

At present, the current landscape of fusion enabling technologies educational and research programs is limited to a few universities with a limited set of courses. In order to grow the field, it will be necessary to initiate new programs and continue and expand existing programs for Fusion Enabling technology ([Carasik et al., 2024 WP](#)).

Some technical elements are unique to a specific fusion concept, some are unique to all fusion concepts, and others are broadly applicable. Technologies that are specific to fusion are:
- plasma-facing components;
- tritium breeder blanket technologies (including neutronics, structural mechanics, and thermal mechanics);
- tritium fuel cycle management (fuel injection, exhaust processing);
- fuel/material injection (cryogenic pellets);
- plasma measurement systems.

At present there are a wealth of competing approaches to realize fusion, but all of the above technologies share broad applicability across fusion concepts. These enabling technologies should be emphasized in educational programs as they provide ample open technical scope, and a versatile education that can be broadly applied ([Carasik et al., 2024 WP](#)).

Even beyond fusion, many of the required enabling technologies are common with other areas of Clean Energy Technology development. These technical areas are particularly attractive to develop workforce development programs around, as graduates of these programs will find employment opportunities across a broad range of fields. This will give their education additional value and enable robust, long-lasting programs of sustained value regardless of the rate of progress towards fusion energy. These technologies are:
- Power conversion (Rankine cycles, Brayton cycles, and advanced Brayton cycles)
- Cryogenic engineering (cooling of the magnets, design of the thermal shield systems)
- Vacuum engineering (exhaust pumping systems, cryogenic vacuum pumping systems)
- Remote handling (Robotics and handling systems for extreme environments)
- Superconducting magnets (for fusion, medical applications, accelerator applications)
- Precision manufacturing (high volume, quality assurance, non-destructive testing)

Regardless of the fusion concept, manufacturing in the private fusion sector will look similar to today's established energy businesses, for example FirstSolar, Tesla, CATL, Tesla, and Vestas. These industries primarily employ mechanical, electrical, instrumentation and controls, chemical, industrial engineers and technicians today, although these industries started with lab-scale prototypes requiring highly specialized advanced degrees. We anticipate that private fusion sector employment will follow a similar trend.

# Educating, Training, and Developing the Fusion Workforce

Fusion-specific degree programs, fusion-related elective courses, and relevant certificate programs are the basis of educational offerings at universities, community colleges, and trade schools. The transition towards sustainable energy solutions has amplified interest in fusion energy, driving educational institutions to offer these specialized programs that prepare students for careers in this field. There is, however, **more work needed to provide the comprehensive curriculum necessary for physics and technology development to fulfill the workforce requirements** of the public and private fusion sector. Here we describe recommendations for the fusion workforce across all educational levels, from K-12 through professional re-training.

Furthermore, to expedite the commercialization of fusion energy within the next decade, the US workforce strategy should focus on strong recruitment and retention efforts targeting a diverse group of professionals at all career levels to meet the growing demand in both the public and private sectors. This need encompasses STEM professionals with degrees in nuclear science, engineering, or related fields, as well as a skilled technical workforce (STW) without bachelor's degrees. Additionally, it is essential to ensure a foundational understanding of fusion for non-STEM professionals, including those in administration, communication, and procurement.

## K-12 Early Education and Outreach to Attract Students

A vital step to train the future workforce is to increase the number of students applying to relevant STEM disciplines. A comprehensive K-12 strategy can raise the awareness and attractiveness of fusion as an education and career option while enhancing public acceptance.

**Develop a comprehensive K-12 public engagement and outreach strategy**

Studies have shown that students who engage with STEM professionals and participate in science-related activities are more likely to pursue STEM degrees: 70% more likely in middle-school students, 85% more likely in high school students (UC Davis, 2024). Such a strategy can include:
- **Outreach and public engagement activities:** Organize and participate in interactive workshops, camps, science fairs, and other related activities where students and the broader public can engage with fusion science through hands-on experiments and demonstrations. Plasma Science Teacher Day equips teachers with the right resources to inspire and educate students about the importance and potential of fusion energy (APS, 2024; [Cruz et al., 2024 WP](#)).
- **Professional visits and guest experts**: Facilitate visits from industry professionals and scientists to provide insights into the fusion industry and its potential impact on clean energy as well as their STEM journey.

- **Student competitions**: Organize student competitions and hackathons to challenge students to solve real-world problems related to fusion energy, fostering critical thinking and innovation. Example: ITER Robots competition (ITER, 2023).
- **Introduce and strengthen fusion in the K-12 curriculum:** Partner with K-12 curriculum providers to incorporate fusion-specific topics into K-12 textbooks and curriculum. Exposure to fusion is particularly valuable at an early stage through examples of applications and problem sets to engage students when selecting their further study/work direction ([Schaffner et al., 2024 WP](#)).
- **Measure impact through metrics**: Conversion rates from outreach activities, information sessions into internships, enrollment in fusion-related courses and degree programs and number of applications for fusion position can indicate effectiveness of the outreach activities supported by surveys identifying the most valuable initiatives.

## The Role of Academia

The concentration of young talent coupled to the intrinsic interdisciplinary nature of universities has the potential to address major needs for a skilled and diverse fusion workforce. Furthermore, some university programs operate small and mid-scale fusion-related devices which provide valuable hands-on experience that can be tailored to a specific fusion concept. University programs are cost-effective and require openness in research, which lowers the barrier for cross-institutional access and improves societal trust in fusion. Finally, university faculty can provide non-conflicted, merit-based guidance in a broad range of fusion-relevant topics beyond science and technology. These include intellectual property, technology transfer, energy market analysis, safety and licensing, customer and end-use adaptation, and societal impact/acceptance.

Beyond education and training, university programs can provide major support for fusion energy through innovation and venture creation, arbitration and consulting, and establishment of social acceptance. According to a survey of the Fusion Industry Association (FIA), most fusion companies are launching out of universities (58% of FIA members), including the first US fusion company (TAE Technologies) and the largest in terms of capital raised (Commonwealth Fusion Systems) ([Wirth et al., 2024 WP](#); Whyte et al., 2023). This points to both the quality of fusion university research and the credibility when attracting investors.

**Launch programs to catalyze faculty positions in fusion engineering and technology (in addition to plasma physics) across diverse academic programs**

Academia's long-term commitment for growing and sustaining fusion programs should be stimulated by government funding. Institutions seeking to sustain/expand existing fusion programs or establish new ones may require assistance in training to teach a new and fast-changing curriculum. This need can be addressed through the establishment of government programs sponsoring new tenure-track faculty positions (research and teaching), training opportunities for faculty, and a diversity of academic programs. These efforts should be aligned with other government goals related to economic development, engaging with minorities, and stimulating geographically balanced competitive research.

- **Increase the number of faculty positions:** The ability for universities to support fusion depends on the number of active fusion programs, the number of faculty per program, the breadth of faculty expertise, and the diversity of academic institutions involved. When compared to other technology-driven industries, including nuclear and aeronautics, there are similar numbers of plasma/fusion university programs is comparable; 57 in plasma/fusion, 40 in fission, and 82 in aero/astro ([Wirth et al., 2024 WP](); Whyte et al., 2023). However, on average, there are only 3 faculty in plasma/fusion per institution, as compared to 19.5 faculty in nuclear and 32 faculty in aero/astro.
- **Support faculty positions in engineering and technology**: With only 15% of faculty with primary focus on materials and/or fusion technology and the remaining 85% being plasma-focused faculty, there is a disbalance in expertise (Whyte et al., 2023). The demand for plasma/fusion-trained professionals increases steadily each year and the desired skill set is shifting from science to engineering and technology. Therefore, when hiring new plasma/fusion faculty, university programs should ensure a balance of expertise, including fusion engineering and technology, in addition to plasma physics.
- **Diversify academic programs:** There is a pressing need to expand and diversify the types of fusion university programs to include master's degrees, bachelor's concentrations, and post-secondary education as well as (short) online certification programs.
- **Support the development of on-campus Infrastructure:** Academic institutions are the ideal locations to host research infrastructure, incubate start-ups, and foster industry collaborations. Academic institutions are often located in population centers with ready access to a diverse talent pool capable of being trained. Only the largest national assets should be sited within the national laboratory complex, where personnel access is restricted and whose sites were chosen based on remoteness. Research infrastructure programs supported by agencies should encourage submissions that work towards fusion-relevant goals and develop the fusion workforce in parallel. Inter-agency partnerships should be exploited to advance key multi-mission infrastructure.

### Establish guest lecturer mobility schemes and development opportunities for subject matter experts from industry and national labs

Subject matter experts from industry, private companies, user facilities, national laboratories, and universities can support training and education activities both at universities, vocational training, and certification programs as part-time lecturers primarily in advanced, niche topics and skills. This can provide mutual benefits to the host and sender organizations creating flexibility in the course topics while ensuring relevant, innovative, advanced and up-to-date knowledge transferred to students. In exchange, sending organizations can benefit from trained candidates to recruit from. The following opportunities should be pursued:

- **"Train the trainer" programs:** Experts without lecturing experience can benefit from programs that provide them with an overview of the latest education techniques and practice over a few days to ensure they can engage students and efficiently transfer their knowhow and experience to them. Such mobility schemes are common in the nuclear fission sector, such as the Candu Owners Group, as well as in national and international fusion summer schools ([Walkden et al., 2024 WP](#)). The same training can broadly benefit other educators.
- **Exposure and new learning opportunities for faculty:** It is recommended to create (funded) programs for visiting faculty members at national laboratories & established universities in order to engage faculty in research activities and build a strong foundation to develop fusion enabling technology research programs at their home institutions.
- **Sabbatical and mobility scheme for professionals**: Professionals working in industry and national labs could also benefit from short-term visits to other organizations, including universities, to learn from best practices and lessons learned at various institutes or be exposed to new topics and research or activity areas. Professionals can also be empowered to deliver curriculum in relevant technical areas.

### Expand and create continuing education, fellowship, and internship opportunities

Student and professional education is a lifetime activity, and scope for additional internship programs at academic institutions exist as described below. It is recommended to create new career development opportunities, including training and internship programs, mentorships, fellowship opportunities. Broadly available fusion-related undergraduate internship opportunities at academic institutions are exceedingly rare, with only the internally supported Plasma and Fusion Undergraduate Research Opportunities (PFUROs) existing. Other opportunities are only for internal applicants or limited to engagements at the national lab and user facilities. Expanding access to existing programs, such as the Fusion Undergraduate Scholars (FUSars) which combine research opportunities with professional development, should be encouraged, as well as NSF Research Experience for Undergraduate (REU) sites focusing on fusion-related topics ([Shulman, 2024 WP](#)). Community Colleges are also excellent locations for supported internships and learning opportunities.

Fellowship opportunities should also be expanded for fusion-related areas. Education at the terminal-masters level is presently more difficult to emphasize due to the common feature that students pay their own tuition. Funding this education level can deliver skilled trainees to support industry more quickly than doctoral education. Agency-funded fusion master's fellowships can also include an internship component, as described later in the Partnerships section.

## Improving the Fusion Curriculum, Software, and Learning Materials

Analysis of the current state of the clean energy workforce reveals a burgeoning interest among students at all levels in fusion energy. However, a significant skills gap persists, necessitating targeted training programs to equip early talent with the requisite skills to meet industry demands ([Makani et al., 2024 WP](#)). Access to introductory and advanced educational material is identified as a key recruitment challenge also in Europe (Fasoli et al., 2024). The diversity of fusion concepts explored in the private sector and the ambitious industry timelines pose several challenges, which are also opportunities for collaborations with academia.

The fusion curriculum needs to cover a breadth of interdisciplinary topics, ranging from basics of plasma physics to fusion engineering and technology development. The educational and training programs should teach a balance between universal skills (that can translate across occupations) and specialized skills, such as work with magnets for magnetic fusion energy (MFE) and work with lasers for inertial fusion energy (IFE). With the growth of the fusion industry, there will also be a substantial need to develop the supply chain for existing fusion components as well as an opportunity for venture creation to deliver first-of-a-kind components not available in the present supply chain. Universities have the appropriate environment to foster interdisciplinary work, innovation through original research, and venture creation needed to address these challenges.

**Develop and implement a national strategy for fusion curriculum leveraging existing resources to close gaps**

To maximize the impact and return on investments into curriculum development, a dedicated funded study is needed to provide a comprehensive picture of the current curriculum landscape in the US. Specifically, identifying a list of plasma/fusion-specific and plasma/fusion relevant courses currently taught and classifying them by institution type and geographical location would allow for both leveraging existing resources and identifying gaps. An initial attempt to provide such information for the state of Virginia was discussed in the white paper about fusion curriculum generated for the NSF-CET meeting ([Mordijck et al., 2024 WP](#)). In addition, similar efforts were already undertaken in Europe and can be used as an example (Fasoli et al., 2024). The EUROfusion consortium in Europe conducts an annual mandatory survey of the academic degree programs, courses and other educational activities and material among the publicly funded research centers and universities in fusion. Additionally, training needs are updated on a yearly basis identifying scarce and vital engineering areas to target funded postgraduate excellence programs. Collaboration with Europe and beyond on a joint international fusion curriculum can help close existing gaps and strengthen both the availability, access and quality of the US and international educational programs.

The outcomes of a national plasma/fusion curriculum study will help federal institutions initiate targeted programs. The standardization of core courses and the spread of such programs can be accomplished through sharing instructional resources across universities. Higher-level fusion curriculum, traditionally covered in PhD programs, should be taught more often in Master's degrees, which are very common targets for graduates who want to step directly into industry.

The assessment and development of the fusion curriculum should engage all stakeholders, including academic institutions, trade unions, and fusion employers ([Gehrig et al., 2024 WP](#); [Wirth et al., 2024 WP](#)). Universities can collaborate with trade unions on the development of highly specialized training programs tailored to the needs of the fusion sector and offer opportunities for hands-on training and apprenticeships. For example, the International Brotherhood of Electrical Workers (IBEW) has partnered with community colleges to offer specialized programs in renewable energy technologies ([Makani et al., 2024 WP](#)). Trade unions also provide a vital connection with potential employers offering help with job placement and career advancement. Further partnership with industry and national labs is necessary to ensure that the fusion curriculum includes both foundational knowledge and emerging fields. Each of these collaborations should employ rigorous assessment tools to evaluate the efficiency of the programs, provide ways to communicate, and adapt to the needs of all stakeholders.

The outcome of this assessment shall provide a structured list of available courses and materials to be visible on the centralized information hub mentioned earlier as well as a prioritized list of gaps for which to create new educational offerings.

**Improve accessibility to education by developing shorter certificate programs and publicly available courses**

Core introductory topics, such as plasma physics, reactor design, materials science, and nuclear engineering, can be offered as micro-credentials and certification programs. Such programs can provide foundational knowledge in a cost-effective and flexible manner, which will help broaden participation in the fusion workforce. These programs are designed to accommodate working professionals, adult learners, and enable students/professionals seeking to upskill or re-skill quickly to enter the fusion sector from neighboring industries.

In collaboration with universities and research centers, Europe is currently developing the Fusion Education and Learning Hub (FuEL) to provide access to academic courses online on-demand, or live through a dedicated Moodle platform. This initiative does not provide accreditation or certification at its early stages, but could provide mutual benefit on the exchange of US and European courses to jointly fill skill gaps, training on scarce (engineering) areas while offering access to the recent breakthroughs and innovations. Furthermore, it can advertise US programs and study/work opportunities to a wider international audience.

**Support the production of engaging, interactive and inclusive educational materials**

Creating or sharing a variety of teaching material (text, graphs or video, interactive project work or capstone projects) can lead to a more interactive and engaging learning experience suitable for different learning preferences and abilities as well as developing comprehensive skills along Bloom's updated taxonomy, shown in Fig 2.

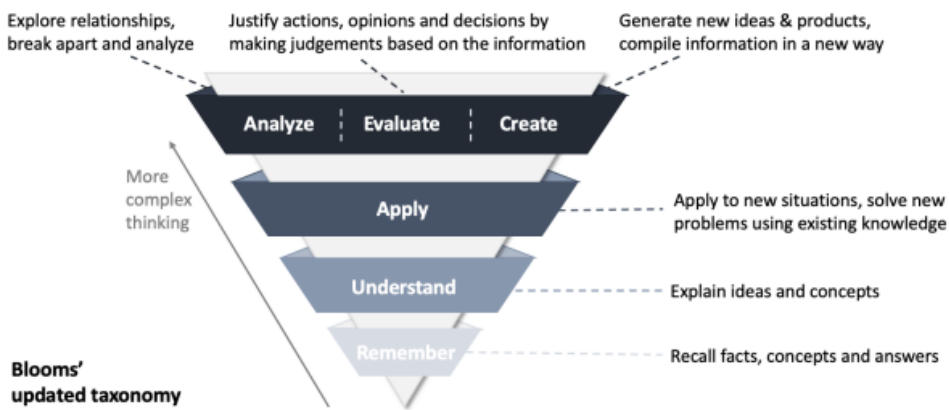

Fig 2: Bloom's updated taxonomy for developing comprehensive skills.

Incorporating a variety of different teaching tools including tools targeting accessibility to different learning capabilities/preferences and physical/mental abilities make learning more inclusive by design. The Universal Design for Learning is a methodology to make learning material and activities inclusive from the start, without the need to modify them afterwards to accommodate student needs (Harvard Graduate School of Education, 2024). These can involve simple steps such as using color palettes suitable for color blindness, using both visual and text elements to support neurodiversity, or recording lectures with subtitles and transcripts for foreign students, beginners and people who like to review/research materials carefully.

Regular modest assessments, spaced repetition methods, and gamification can further ensure retaining core learning points, increased student satisfaction, and prolong engagement to complete/enjoy the course. Innovative teaching tools, such as the use of AI tools or virtual reality (VR) simulations, allow the students to better engage with the learning material and visualize complex fusion processes making learning more accessible and captivating. One example is the Imperial College IDEA Lab (Imperial College Business School, 2024). By soliciting regular feedback from students, lecturers can further identify barriers and opportunities to optimize their teaching to benefit students.

### Encourage open-source ecosystem development for fusion hardware and software

Workforce development is most readily achieved when software tools are easily accessible and available for pedagogical use. Open-source ecosystems, for both hardware and software development, offer the most effective means to facilitate this engagement. These types of ecosystems have found great success in other areas (e.g., AI/ML, robotics, rocketry) supporting concurrent R&D and training. The benefits of open-source ecosystems include:
- Training and education become closely linked with industry tools and methodologies, enabling innovative teaching methods that support DEI goals at all education levels.
- Lower barrier to entry (such as cost, training) for new stakeholders.
- A quick transfer of advancements across different stakeholders. Insights obtained from the industry can immediately contribute to research and training at Academia and National Labs.
- Fusion systems will integrate many existing and emerging technologies from other fields with open-source ecosystems, and similarly, other technologies will also benefit from cooperative development.

It is important to prioritize the development of structures and programs that support and steer this ecosystem. This should involve input from all members of the community, with a focus on ensuring fair and inclusive participation and access ([Hansen et al., 2024 WP](#)).

### Community Call-to-Action: Encourage program elements to capture tacit knowledge through codification into learning materials and direct transfer from experts

Fusion is still in its Research and Development (R&D) phase with new discoveries, niche expertise and decade-long experience existing in the minds of a few key experts, as unwritten tacit knowledge. This is particularly true in the design, commissioning and operation of fusion devices where repeating mistakes can lead to months/years of delays in construction and commissioning of new devices. Examples exist of programs to convert tacit, unwritten knowledge into explicit knowledge (Belonohy, 2024; Rogers & Ryschkewitsch, 2008). Performers should encourage activities involving various stakeholders to improve the capture and sharing of tacit knowledge between institutions. This goes beyond physics research to scarce skill areas, with a goal of building a comprehensive knowledge hub which is broadly available to all performers. These efforts should be complemented by (AI-driven) extensive tools to enable easy access to the generated knowledge and identify the right experts to speak to.

# Expanding the Diversity of Academic Institutions

Establishing new academic programs is crucial to expand the workforce, ensure geographically balanced fusion education, and engage with underserved communities and underrepresented groups. This can be best achieved through engagement with a diversity of institutions, including public and private universities, primarily undergraduate institutions (PUIs), minority serving institutions (MSIs), and community colleges. Good candidates for fusion programs can be institutions with fusion-relevant existing programs, such as materials science, engineering, and physics. Other good candidates for new fusion programs may exist at universities located near fusion facilities or national laboratories with strong fusion fusion. Small universities seeking to build out interdisciplinary science and engineering programs for students from underserved or minority backgrounds are excellent candidates for the establishment of fusion bridge programs in partnerships with bigger research universities.

PUIs, where teaching is often the primary objective, can be instrumental in fusion curriculum development, pedagogical approaches, effective student engagement, and assessment ([Schaffner et al., 2024 WP](#)). In addition, faculty at PUIs can be an important point of contact between fusion research institutions and the pool of potential workforce members, particularly given the relatively few centers of fusion study across the country and the world. Last, but certainly not least, schools wishing to promote fusion in general but lacking the expertise are ideal candidates for fusion program development. Some of these institutions may also benefit from non-technical but relevant curricula on activities which promote energy science through law, economic studies, policy, history, educator-focused, communication, etc. An example of this gaining traction is teaching fusion history and using this as a valuable knowledge transfer tool.

Currently, NSF offers multiple programs that support hiring of faculty in needed areas, such as Building Research Capacity of New Faculty in Biology and Faculty Development in the Space Sciences. In addition, the Faculty Early Career Development Program has been an excellent avenue for identifying and sustaining early career talent. Building on these examples, the fusion workforce would greatly benefit from the establishment of a dedicated program focused on stimulating faculty hires in fusion and fusion-adjacent fields in addition to traditional plasma physics.

**Community Call-to-Action: Fusion stakeholders should facilitate the establishment of regional plasma/fusion centers of excellence**

A long-term goal of expanding university fusion programs should be the eventual establishment of plasma/fusion departments. An initial step in that direction is funding plasma/fusion centers across departments or cross-college institutes. This would allow the flexibility to offer interdisciplinary education, utilize curriculum already available in fusion-relevant fields, and attract students and trainees with diverse backgrounds and interests.

Academic institutions can partner with each other to expand curricular offerings as well. The UK established its government-funded Engineering and Physical Sciences Research Council (EPSRC) Fusion Centre for Doctoral Training (CDT) in Fusion Power, which involves six leading research-intensive universities, 20+ private companies, national and international labs and government agencies (Fusion CDT, n.d.). Collaborative efforts between several universities and/or industry partners can lead to the creation of joint programs that provide students with extensive curriculum and hands-on experience in fusion technology. An example is University of Wisconsin's partnership with national laboratories (University of Wisconsin-Madison, 2024).

All stakeholders should work towards the establishment of new fusion programs in diverse academic institutions by supporting new faculty positions (both research and teaching), training current professors through webinars and workshops, and establishing faculty/scientist exchange programs. This includes establishing a dedicated agency-funded program focused on stimulating faculty hires in fusion and fusion-adjacent fields. Major fusion academic institutions and private companies should reach out to schools within their vicinity and aim to spur the growth of fusion relevant programs through financial and technical support.

# Domestic and International Partnerships

Institutions with profound experience and capabilities in the fusion ecosystem are already in existence. This section will discuss how these existing resources can be leveraged.

**Develop partnerships with existing public laboratories, private sector and other relevant tech sectors to increase retention and recruitment**

Laboratories in fusion have long served the nation as training centers largely funded by government programs. Collaboration with academia and the private sector is required to meet the necessary workforce growth to fulfill the private sector needs. Since the private sector is a main stakeholder it should support this growth through fusion faculty endowments and direct support for students and postdocs through employment opportunities, fellowships and stipends. These activities can also be enabled by government-sponsored public-private partnerships. Many of these efforts have similarities on the private and public side, which should be exploited as described below:

- **Knowledge transfer:** Various workforce development programs exist at each public laboratory, academic institution, and private companies. Best practices, lessons learned, and training materials should be widely shared, continuously improved, and coordinated at the national level.
- **Recruiting a diverse workforce:** Collaborative efforts between non-educational and educational institutions greatly benefit recruitment. Students and early career professionals with exposure to industry professionals and real-world applications are 60% more likely to pursue careers in those fields (UC Davis, 2024). Additionally, targeted recruitment and onboarding programs for mid-career professionals have been shown to reduce turnover rates by up to 40%, indicating successful career transitions (UC Davis, 2024). Informational sessions, facility and company visits, workshops and seminars at universities, community colleges and trade schools could be used to target students in disciplines beyond nuclear science by highlighting how a new job entrant's skills can be applied to fusion research and the opportunities available in the industry. University and college student networks (newsletters, social media) can be leveraged to disseminate information about fusion sector opportunities. Similarly, information sessions targeting professionals considering a career change highlighting transferable skills, growth opportunities and exciting challenges in the fusion industry should also be hosted. Also, comprehensive onboarding programs for mid-career professionals entering the fusion field including mentorship, training modules, and job rotations could be developed.

- **Regional Clusters:** Public laboratories and private companies should continue to expand and build relationships with academic institutions, particularly local ones that could lead to a diverse workforce. Establishing regional clusters is an established method to accelerate growth. It requires the presence of private companies, university programs, and supporting national laboratories or user facilities. Six regions in the US currently fulfill these requirements and could therefore become fusion clusters: California, Massachusetts, New York and New Jersey, Tennessee, Washington, and Wisconsin ([Deshpande, 2024 WP](#)).
- **Internships and Apprenticeships:** Internships outside of academia are integral for students to gain experience as well as explore various career paths and topics to decide on their future career. Most internships are through the government funded programs SULI and CCI programs. Industrial internships, particularly for non-scientist positions, could help meet the workforce needs. Vocational skills are essential to support research and development activities. In Europe, the number of technicians halves every 10 years. Skills shortages and lack of experience can lead to long delays on the construction and commissioning phases of large projects as well as delay operation in the scientific exploitation phase. In partnership with the U.S. Department of Labor, the Princeton Plasma Physics Laboratory (PPPL) has recently started a 4-year registered apprenticeship program for technicians open to anyone with a high school degree or equivalent to develop a high-quality career path into the fusion workforce at their laboratory (Huber, 2024). The United Kingdom Atomic Energy Authority (UKAEA) established the renowned Oxfordshire Advanced Skills (OAS) center at the Culham Campus, which provides award-winning apprentice training for vocational skills for fusion and high-tech industries in the region (OAS, n.d.).
- **Retention:** Support existing fusion professionals through continuous training, leadership and mentorship opportunities, sabbatical and career mobility schemes to increase exposure to a variety of projects. Additionally, create mechanisms to provide foundational fusion knowledge to non-STEM professionals in support roles.

Joint educational and training programs can be created through government agency funding. Furthermore, existing networks in scientific exploitation and technology can be extended to other areas such as administration, procurement, training and education through the American Physical Society. LaserNetUS, a network of high-power laser facilities in North America funded by DOE's Office of Fusion Energy Sciences, has a strong workforce development and training component as an integral part of its mission ([Wei & Obst-Huebl, 2024 WP](#)). These networks or communities of practice can involve other large-scale research organizations in physics, engineering or other STEM fields (e.g., aerospace, particle physics, biology) similar to EIROforum, which involves the 8 largest research centers in Europe (EIROForum, 2024).

### Establish collaborative projects between US institutions and international partners on education and workforce development

Extension of existing collaborations with international partners on scientific exploitation and technology should be expanded to education and workforce development. Examples include:

- **International Atomic Energy Agency (IAEA)**: IAEA is renowned for its support of the international nuclear fission activities. The agency is currently expanding into fusion with dedicated effort on workforce development, training and knowledge management. Its recent *International Conference on Nuclear Knowledge Management and Human Resources Development 2024* brought together experts working in the nuclear fission and fusion sector with strong effort on training and education methodologies and best practices (IAEA, 2024). Additionally, its IAEA CONNECT education platform is expanding to include new content for fusion education (IAEA, n.d). Fusion can leverage and build on the existing IAEA connections with international partners (e.g. Global South) to develop collaborations with international partners.
- **Global South:** Although research and development in fusion energy has benefited from international collaborations (e.g., ITER, IFMIF), relatively few fusion programs exist in the Global South (IPP, n.d.). This is an energy justice issue, as countries in the Global South have a) increasing energy needs, b) unique energy needs (e.g., district heating, desalination), and c) have contributed the least to climate change. There are a number of groups actively engaged in fusion and plasma research in Latin America (Argentina, Brazil, Chile, Costa Rica, Mexico), North Africa and the Middle East (Arab Fusion Energy Initiative, Egypt, Libya, Morocco, Tunisia) and Asia (Malaysia, Thailand). We advocate for partnerships between countries in both regions, to prevent geopolitical tensions over the unequal distribution of elements and specialized materials needed to support technological advances in fusion energy ([Shah et al., 2024 WP](#)). Providing platforms for the exchange of ideas, creating international programs in education and outreach, and implementation of projects to address current economic barriers to progress would increase access for partners located in the Global South. Such collaborations may be supported by NSF MOUs led by the Office of International Science and Engineering and Technology and the Directorate for Technology, Innovation and Partnerships. Additionally, there is a need for projects that focus on promoting awareness of and building trust in fusion energy in the Global South. Long term, the establishment of a supply chain between regions and the involvement of governments in the Global South will provide opportunities for all.
- **UK national fusion program**: As part of the UK Fusion Futures Programme, the UK plans to train 2200 people in the next 5 years, to meet the growing needs of the UK fusion sector ([Walkden et al., 2024 WP](#)). The Fusion Opportunities in Skills, Training, Education and Research (FOSTER) works with business and universities to increase the number of apprenticeships, graduate programs, and postdoctoral training positions in the UK. There are existing strong ties with the UK's national laboratory (UKAEA) that can be expanded to include training and education efforts.
- **EUROfusion consortium in Europe**: In Europe, the research and development program is coordinated jointly by the EUROfusion Consortium involving 197 fusion institutes including 100 universities in the European Union funded by the European Commission with partners in the UK, Switzerland, and Ukraine. A strong element of the program is Training and Education. EUROfusion supports outreach activities, school teachers, internships and participation in educational events of graduate students through FuseNet ([Cruz et al., 2024 WP](#)); 1000+ MSc and PhD students annually contributing to their salaries and missions through the EUROfusion member institutes; and 30+ 2-year post-graduate scholarships annually in research and engineering aligned with scarce engineering competencies.
- **Japan and South Korea:** Japan is recently establishing a new national strategy for fusion energy, and ample opportunities for technical and educational collaboration exist. The Broader Approach (BA) Agreement has already demonstrated a valuable partnership between Japan and Europe (European Commission, 2020). South Korea has a longstanding partnership with US scientists in magnetic confinement fusion research, which could be expanded to education.

# Workforce Needs and Scaling

The final makeup of the fusion workforce still possesses significant uncertainties, as does its eventual rate of growth. This is in part due to the wide range of technical approaches being presently pursued towards fusion, each with their own specific workforce needs. The most successful concept is still not determined, requiring multiple workforce options to be explored in parallel. To address these uncertainties the following recommendations and calls to action are presented.

**Launch a formal study evaluating the distribution, needs and challenges of the current and future US fusion workforce**

The private sector fusion workforce is expected to grow substantially, with projections indicating a size of over 25,000 workers by 2035 ([Deshpande & Segal, 2024 WP](#)). This represents a scaling factor of more than 18 in the next decade, with an average annual growth rate exceeding 30%. However, the exact makeup of the workforce skills for essential staff at all levels is not fully understood in the US. The ratio between STEM (Science, Technology, Engineering, and Math) professionals; science and engineering related occupations such as project management and policy; and skilled technical workers, is not known. It is expected the future fusion workforce will be significantly broadened from its present-day doctoral graduate base. Europe recently conducted a workforce assessment of the publicly funded research institutes identifying needs, challenges to establish their workforce development strategy (Fasoli et al., 2024). Effective data collective of current diversity levels would allow analysis and actionable next steps for understanding what holds us back from an inclusive workforce (Burnett et al., 2022; Kaplowitz & Laroche, 2020).

**Document the level and trajectory of accessibility and diversity in the fusion workforce**

Supporting the growth of the fusion industry will require an investment in understanding and addressing the long-standing challenge to recruit and retain a diverse workforce. The fusion workforce is estimated to be 12-23% women, 65-80% White, 8-27% Asian, 4-9% Hispanic/Latinx, 1-3% Black or African American, and 0.1-0.7% Native American or Alaska Native (FIA, 2023; May, 2022; The Fusion Cluster, 2023; W+IPP, 2024; Zippia Research Team, 2024). We recommend that new activities for engagement include targeted recruitment strategies to reach a wide range of groups, including women and others who are currently under-recognized in the fusion industry. In addition to goals associated with increasing diversity in race, ethnicity, and gender identity, we advocate for activities that focus on increasing diversity in academic paths, area of expertise, disability, and socioeconomic background ([Coté et al., 2024 WP](#)). The latter can be a major challenge to equitable education access, which results in massive loss of talent in the workforce. As the fusion industry continues to develop, comprehensive data on workforce diversity should be regularly collected and reported. This data should include the categories often overlooked in current workforce reports and consider how intersectionality may impact fusion-related career pathways. Additionally, tools like the "Diamond Model" and "Queryable Equity Database" can be used to support systemic change at institutions and align decision-making with the experiences and needs of BIPOC, LGBTQIA+, gender minorities, socio-economically challenged individuals and other groups that have been historically excluded from STEM fields (DPP DEI OCC, n.d.; [James & Alfread, 2024 WP](#)). New studies are needed to a) document the variety of fusion-focused workforce programs that currently exist, b) develop program assessment tools, c) study the impacts of current programs that have been successfully implemented, and d) identify gaps that can be addressed through targeted workforce projects ([Coté et al., 2024 WP](#)).

**Community Call-to-Action: Support the centralized nationwide online listing of educational resources, training opportunities, and job opportunities**

To attract students and mid-career professionals to join the fusion workforce as well as retain and train the current workforce, it is vital to support, fund and ensure visibility of a centralized hub for fusion resources. Imagined as a prominent website (such as [https://usfusionenergy.org)](https://usfusionenergy.org), this hub will be actively maintained and list national and international resources for educational, training, and job opportunities. Increasing support to tools such as these will facilitate the engagement with a broad workforce and also expedite filling available opportunities. Existing scientific societies could also partner to deliver this comprehensive service.

# Community Benefits and Inclusion

New studies and initiatives should increase diversity in the fusion workforce, improve working conditions, improve collaborations across institution types, and provide global access to new scientific instrumentation, technology, and infrastructure.

**Community Call-to-Action: Strengthen the fusion community through collaboration, equity, and inclusion**

Collaborations among all stakeholders, including companies, federal agencies, academia, and national laboratories, are needed to develop and sustain the next phase of the fusion energy workforce (Carter et al., 2020; Whyte et al., 2023). There must be a widespread acknowledgment within the fusion community that having a diverse and equitable workforce is a priority. This requires actionable, effective techniques for fostering inclusion, which should be regarded as essential for the community and industry's advancement. Defining and measuring goals for diversity, equity, inclusive, and accessibility (DEIA) over the course of a career is critical to making ongoing improvements. Without clear, quantifiable targets, it is challenging to gauge progress or understand the effectiveness of DEIA initiatives.

With respect to gender identity, ethnicity, race, disability status, and parental educational attainment, Minority Serving Institutions (MSIs) and community colleges are collectively more diverse than students attending other schools in the US. Thus, a major step in diversifying the fusion workforce is the establishment of meaningful partnerships with diverse types of institutions, including MSIs, community colleges, and PUIs. To expand the reach of existing internships (e.g., [CCI](#), [REU](#), [SULI](#)) and fellowships ([GEM](#)), the NSF and other funding agencies could support the establishment of national programs for college/university students, recent graduates, and other trainees to spend a semester (or longer) at a national laboratory to work on technical or research projects in fusion energy.

Despite the number of existing opportunities, short and long-term equity gaps persist when considering the current role of MSIs in diversifying the fusion workforce. Those gaps span the areas of research administrative capabilities, number of research projects awards, field-specific training programs, representation in professional societies, inclusion in public engagement activities, and sustained support for historically marginalized undergraduate and graduate students, postdocs, and faculty ([James & Alfread, 2024 WP](#)). To address national concerns over the geographic inequities in access to high-quality STEM education and careers, we propose increased engagement with MSIs (Honey et al., 2020). The US has more than 800 MSIs, which are distributed throughout the country (Deitz & Henke, 2023). Previous and existing efforts to diversify the fusion workforce have initiated partnerships between MSIs and institutions with substantial experience and resources through programs like the DOE RENEW initiative and NSF's Tribal Colleges and Universities Program, HBCUs Undergraduate Program, and IUSE: HSIs. We advocate for an NSF MSI program focused on plasma physics and fusion energy, to increase diversity in this field. The states eligible for EPSCoR funding are those that receive less federal funding for STEM R&D, and 28 US states currently fall in that category. Currently, there is one NSF EPSCoR project focused on the development of plasma technology in Alabama; Future Technologies and enabling Plasma Processes (FTPP). This project engages 9 universities in Alabama, 4 of which are HBCUs. Lessons learned from this project can serve as a basis for future similar programs. We advocate for a collaboration between the NSF EPSCoR program, MSI-focused programs, and the programs traditionally supporting plasma- and fusion-relevant research, to advance NSF's goal of building bridges across communities. Modeled after the American Chemical Society (ACS) *Green Chemistry Institute*, a "Fusion Institute" could provide infrastructure for the community, conferences, educational resources, and materials to be used during public engagement events (ACS, n.d.).

**Community Call-to-Action: Support efforts that improve working conditions for the existing fusion workforce**

In the U.S., there are many efforts to recruit new people into STEM disciplines, but it is important to take actions to promote retention for those already working in the field. As education and career stage increases, the diversity of STEM professionals and students decreases (Kali Pal et al., 2023; Porter & Ivie, 2019). In physics, women report receiving less funding, awards, equipment, laboratory space, invitations to present their work, offers to collaborate on projects, and opportunities to serve as journal editors (Ivie et al., 2013; Porter & Ivie, 2019; Sarsons et al., 2021; W+IPP, 2024). Black and Hispanic/Latinx students and professionals in STEM have less access to mentoring, discipline-specific networking activities, and professional development opportunities (Abrica et al., 2022; Cervantes, 2021; Morton, 2021). Students from low-income backgrounds have less access to activities, coursework, and paid professional development opportunities that would support their entry and retention in STEM disciplines (e.g., Cooper et al., 2021; Coté, 2023; Diemer et al., 2020; Nature, 2016; Pierszalowski et al., 2021). These factors suggest that women, people of color, and individuals from other groups under recognized in fusion and plasma physics will be supported to stay in these fields when there is an equitable distribution of professional opportunities and resources within departments or institutions. Programs and initiatives that provide women, people of color, and people with disabilities with academic or professional mentoring, formal recognition of their technical achievements, leadership opportunities, and access to networking and collaborations with others in their discipline (Abrica et al., 2022; Castro et al., 2024; Jain et al., 2020; Morton, 2021; Singleton et al., 2021). Projects that include a commitment to ethical standards, access to mentoring, training related to inclusion and psychological safety, activities to reveal the "hidden curriculum" about success in fusion/STEM environments, policies to prevent toxic environments, and oversight to enforce these policies will improve and maintain healthy and supportive environments in the fusion industry (Coté et al., 2024 WP; Marder et al., 2024).

**Support community trust and public engagement of fusion energy and its benefits using Community Benefit Planning**

The successful adoption of fusion energy relies on the establishment of trust through a clear multi-way communication among all stakeholders, including communities served and underserved, government, the private sector, Tribal Entities, labor unions, and others. These relationships will be essential for building social acceptance and obtaining a "social license" for fusion energy (Hoedl, 2022, 2023). An important initial step in building trust is battling misinformation. The fusion community would benefit from engagement with existing programs, such as the APS's Science Trust Project (STP), a model for community engagement (APS, n.d.). The STP seeks to empower members and science colleagues to engage in productive conversations about science with skeptical members of their communities, thereby building a community of practice to address misinformation. This is accomplished by (1) aiding APS members in initiating and sustaining productive dialogues around science, shifting from correction mindset to a connection mindset; (2) building a community of practice actively engaged in addressing misinformation to build trust in science; and (3) training members to facilitate STP workshops with their communities, expanding STP reach and impact.

Another crucial activity is the development of a **Community Benefit Plan** that would distribute benefits equitably and lead to broadly shared prosperity ([Keane et al., 2024 WP](#)). In the past few years, the federal government has strongly encouraged federal agencies to consider how programs related to climate, clean energy, affordable and sustainable housing, and other investments affect disadvantaged communities that are marginalized by underinvestment and overburdened by pollution (The White House, n.d.). Especially relevant for federal programs with a strong focus on local impact (e.g., EPSCoR programs), the fusion industry and the broader fusion community should develop plans describing how the significant benefits from fusion will flow to underserved communities. It is important to recognize that urban and rural disadvantaged communities have different issues. Particular attention should be given to communities whose income relies on modes of energy production that would be displaced by fusion. For these communities retraining and other assistance will be necessary. This is important both ethically and politically. To aid these processes, a) federal agencies that fund fusion and fusion-adjacent programs should a) encourage broader impact project elements that promote environmental and energy justice, and b)  fusion industries, in collaboration with federal agencies, professional societies, and the broader community should dedicate resources to building trust and battling misinformation through the development of a Community Benefit Plan for fusion ([Keane et al., 2024 WP](#)).

# Conclusion

Significant opportunities to expand effort to deliver a diverse and capable workforce to support the development of fusion energy technologies. The aforementioned recommendations provide an actionable template for government agencies and the community at large. By working together and leveraging the strengths of each performer class, the creation of this needed workforce can be accelerated. The rapid onset of climate change demands action, and the community is ready to rise to the challenge.

# Acknowledgement

This work was funded by the National Science Foundation through a Division of Physics Grant Award Number 2346410. We are grateful to the Hampton University staff who were integral to the success of the Workforce Accelerator for Fusion Energy Development conference. C. Paz-Soldan. was responsible for supervision and project administration. C. Lowe, C. Paz-Soldan, and T. Carter. were responsible for funding acquisition. C. Lowe, C. Paz-Soldan, D.A. Schaffner, E. Belonohy, E. Kostadinova, L.E. Coté, and T. Carter were responsible for conceptualization, curation, formal analysis, investigation, original draft writing, review, and editing. B. Makani, J. Deshpande, K. Kelly, K.E. Thome, S. de Clark., S.L. Sharma and V. Kruse were responsible for conceptualization, curation, formal analysis, and investigation. Many of the findings and recommendations summarized in this report were outlined in detail in the white papers submitted for the conference, and we are grateful to all of these authors for their contributions.